# Application of Traditional Vaccine Development Strategies to SARS-CoV-2

This manuscript (permalink) was automatically generated from greenelab/covid19-review@e166773 on January 20, 2023. It is also available as a PDF. It represents one section of a larger evolving review on SARS-CoV-2 and COVID-19 available at https://greenelab.github.io/covid19-review/

**This in progress manuscript is not intended for the general public.** This is a review paper that is authored by scientists for an audience of scientists to discuss research that is in progress. If you are interested in guidelines on testing, therapies, or other issues related to your health, you should not use this document. Instead, you should collect information from your local health department, the CDC's guidance, or your own government.

# Authors

- **Halie M. Rando** 0000-0001-7688-1770 rando2 tamefoxtime  Department of Systems Pharmacology and Translational Therapeutics, University of Pennsylvania, Philadelphia, Pennsylvania, United States of America; Department of Biochemistry and Molecular Genetics, University of Colorado Anschutz School of Medicine, Aurora, Colorado, United States of America; Center for Health AI, University of Colorado Anschutz School of Medicine, Aurora, Colorado, United States of America; Department of Biomedical Informatics, University of Colorado Anschutz School of Medicine, Aurora, Colorado, United States of America · Funded by the Gordon and Betty Moore Foundation (GBMF 4552); the National Human Genome Research Institute (R01 HG010067)

- **Ronan Lordan** 0000-0001-9668-3368 RLordan el_ronan  Institute for Translational Medicine and Therapeutics, Perelman School of Medicine, University of Pennsylvania, Philadelphia, PA 19104-5158, USA; Department of Medicine, Perelman School of Medicine, University of Pennsylvania, Philadelphia, PA 19104, USA; Department of Systems Pharmacology and Translational Therapeutics, Perelman School of Medicine, University of Pennsylvania; Philadelphia, PA 19104, USA

- **Alexandra J. Lee** 0000-0002-0208-3730 ajlee21  Department of Systems Pharmacology and Translational Therapeutics, University of Pennsylvania, Philadelphia, Pennsylvania, United States of America · Funded by the Gordon and Betty Moore Foundation (GBMF 4552)

- **Amruta Naik** 0000-0003-0673-2643 NAIKA86  Children's Hospital of Philadelphia, Philadelphia, PA, United States of America

- **Nils Wellhausen** 0000-0001-8955-7582 nilswellhausen · Department of Systems Pharmacology and Translational Therapeutics, University of Pennsylvania, Philadelphia, Pennsylvania, United States of America

- **Elizabeth Sell** 0000-0002-9658-1107 esell17 · Perelman School of Medicine, University of Pennsylvania, Philadelphia, Pennsylvania, United States of America

- **Likhitha Kolla** 0000-0002-1169-906X likhithakolla lkolla2018 · Perelman School of Medicine, University of Pennsylvania, Philadelphia, Pennsylvania, United States of America · Funded by NIH Medical Scientist Training Program T32 GM07170

- **COVID-19 Review Consortium**

- **Anthony Gitter** 0000-0002-5324-9833 agitter anthonygitter · Department of Biostatistics and Medical Informatics, University of Wisconsin-Madison, Madison, Wisconsin, United States of America; Morgridge Institute for Research, Madison, Wisconsin, United States of America · Funded by John W. and Jeanne M. Rowe Center for Research in Virology

- **Casey S. Greene** 0000-0001-8713-9213 cgreene GreeneScientist · Department of Systems Pharmacology and Translational Therapeutics, University of Pennsylvania, Philadelphia, Pennsylvania, United States of America; Childhood Cancer Data Lab, Alex's Lemonade Stand Foundation, Philadelphia, Pennsylvania, United States of America; Department of Biochemistry and Molecular Genetics, University of Colorado Anschutz School of Medicine, Aurora, Colorado, United States of America; Center for Health AI, University of Colorado Anschutz School of Medicine, Aurora, Colorado, United States of America; Department of Biomedical Informatics, University of Colorado Anschutz School of Medicine, Aurora, Colorado, United States of America · Funded by the Gordon and Betty Moore Foundation (GBMF 4552); the National Human Genome Research Institute (R01 HG010067)

**COVID-19 Review Consortium:** Vikas Bansal, John P. Barton, Simina M. Boca, Joel D Boerckel, Christian Brueffer, James Brian Byrd, Stephen Capone, Shikta Das, Anna Ada Dattoli, John J. Dziak, Jeffrey M. Field, Soumita Ghosh, Anthony Gitter, Rishi Raj Goel, Casey S. Greene, Marouen Ben Guebila, Daniel S. Himmelstein, Fengling Hu, Nafisa M. Jadavji, Jeremy P. Kamil, Sergey Knyazev, Likhitha Kolla, Alexandra J. Lee, Ronan Lordan, Tiago Lubiana, Temitayo Lukan, Adam L. MacLean, David Mai, Serghei Mangul, David Manheim, Lucy D'Agostino McGowan, Jesse G. Meyer, Ariel I. Mundo, Amruta Naik, YoSon Park, Dimitri Perrin, Yanjun Qi, Diane N. Rafizadeh, Bharath Ramsundar, Halie M. Rando, Sandipan Ray, Michael P. Robson, Vincent Rubinetti, Elizabeth Sell, Lamonica Shinholster, Ashwin N. Skelly, Yuchen Sun, Yusha Sun, Gregory L Szeto, Ryan Velazquez, Jinhui Wang, Nils Wellhausen

Authors with similar contributions are ordered alphabetically.

## 0.1 Abstract

Over the past 150 years, vaccines have revolutionized the relationship between people and disease. During the COVID-19 pandemic, technologies such as mRNA vaccines have received attention due to their novelty and successes. However, more traditional vaccine development platforms have also yielded important tools in the worldwide fight against the SARS-CoV-2 virus.

A variety of approaches have been used to develop COVID-19 vaccines that are now authorized for use in countries around the world. In this review, we highlight strategies that focus on the viral capsid and outwards, rather than on the nucleic acids inside. These approaches fall into two broad categories: whole-virus vaccines and subunit vaccines. Whole-virus vaccines use the virus itself, either in an inactivated or attenuated state. Subunit vaccines contain instead an isolated, immunogenic component of the virus. Here, we highlight vaccine candidates that apply these approaches against SARS-CoV-2 in different ways. In a companion manuscript, we review the more recent and novel development of nucleic-acid based vaccine technologies.

We further consider the role that these COVID-19 vaccine development programs have played in prophylaxis at the global scale. Well-established vaccine technologies have proved especially important to making vaccines accessible in low- and middle-income countries. Vaccine development programs that use established platforms have been undertaken in a much wider range of countries than those using nucleic-acid-based technologies, which have been led by wealthy Western countries. Therefore, these vaccine platforms, though less novel from a biotechnological standpoint, have proven to be extremely important to the management of SARS-CoV-2.

## 0.2  Importance

As of January 18, 2023, there have been over 667,918,207 SARS-CoV-2 cases, and the virus has taken the lives of at least 6,729,898 people globally ([1](#)). The development, production, and distribution of vaccines is imperative to saving lives, preventing illness, and reducing the economic and social burdens caused by the COVID-19 pandemic. Vaccines that use cutting-edge biotechnology have played an important role in mitigating the effects of SARS-CoV-2. However, more traditional methods of vaccine development that were refined throughout the twentieth century have been especially critical to increasing vaccine access worldwide. Effective deployment is necessary to reducing the susceptibility of the world's population, which is especially important in light of emerging variants. In this review, we discuss the safety, immunogenicity, and distribution of vaccines developed using established technologies. In a separate review, we describe the vaccines developed using nucleic acid-based vaccine platforms. From the current literature, it is clear that the well-established vaccine technologies are also highly effective against SARS-CoV-2 and are being used to address the challenges of COVID-19 globally, including in low- and middle-income countries. This worldwide approach is critical for reducing the devastating impact of SARS-CoV-2.

## 0.3  Introduction

The development of vaccines is widely considered one of the most important medical advances in human history. Over the past 150 years, several approaches to vaccination have been developed and refined ([2](#)). The COVID-19 pandemic has produced unusual circumstances compared to past health crises, leading to differences in vaccine development strategies. One way in which the COVID-19 pandemic differs from prior global health crises is that the SARS-CoV-2 viral genome was sequenced, assembled, and released very early in the course of the pandemic (January 2020). This genomic information has informed the biomedical response to this novel pathogen across several dimensions (3, 4). All the same, vaccines have been developed since long before the concept of a virus or a viral genome was known, and as early as September 2020, there were over 180 vaccine candidates against SARS-CoV-2 in development, many of which employed more traditional vaccine technologies (5). However, public attention in the United States and elsewhere has largely focused on vaccine development platforms that use new technologies, especially mRNA vaccines. We review vaccine technologies used for SARS-CoV-2 in two parts: here, the application of established vaccine development platforms to SARS-CoV-2, and separately, novel nucleic acid-based approaches (6).

Understanding vaccine development programs that are using well-established technologies is important for a global perspective on COVID-19. As of December 2, 2022, 50 SARS-CoV-2 vaccines have been approved for use in at least one country (7). A resource tracking the distribution of 28 vaccines indicates that, as of January 18, 2023, 13.0 billion doses have been administered across 223 countries (8). Many of these vaccines use platforms that do not require information about the viral genome, with 20 developed using subunit and 11 using whole-virus approaches (7). The types of vaccines available varies widely throughout the world, as the process of developing and deploying a vaccine is complex and often requires coordination between government, industry, academia, and philanthropic entities (9).

Another difference between prior global health crises and the COVID-19 pandemic is the way that vaccines are evaluated. A vaccine's success is often discussed in terms of vaccine efficacy (VE), which describes the protection conferred during clinical trials (10). The real-world protection offered by a vaccine is referred to as its effectiveness (10). Additionally, protection can mean different things in different contexts. In general, the goal of a vaccine is to prevent disease, especially severe disease, rather than infection itself. As a proxy for VE, vaccine developers often test their candidates for serum neutralizing activity, which has been proposed as a biomarker for adaptive immunity in other respiratory illnesses (11). The duration and intensity of the COVID-19 pandemic has made it possible to test multiple vaccines in phase III trials, where the effect of the vaccines on a cohort's likelihood of contracting SARS-CoV-2 can be evaluated, whereas this has not always been feasible for other infectious diseases. In some cases (e.g., SARS), the pathogen has been controlled before vaccine candidates were available, while in others (e.g., MERS), the scale of the epidemic has been smaller. Vaccine development is traditionally a slow process, but the urgency of the COVID-19 pandemic created an atypical vaccine development ecosystem where fast development and production was prioritized. Estimates of VE have been released for many vaccine candidates across a number of technology types based on phase III trial data.

However, efficacy is not a static value, and both trial efficacy and real-world effectiveness can vary across location and over time. Shifts in effectiveness in particular have been an especially heightened topic of concern since late 2021 given the potential for variants of SARS-CoV-2 to influence VE. Due to viral evolution, vaccine developers are in an arms race with a pathogen that benefits from mutations that reduce its susceptibility to adaptive immunity. The evolution of several variants of concern (VOC) presents significant challenges for vaccines developed based on the index strain identified in Wuhan in late 2019. We discuss these variants in depth elsewhere ([12]). To date, the most significant VOC identified are Alpha (2020), Beta (2020), Gamma (2020), Delta (2021), and Omicron (2021), with various subvariants of Omicron being the most recently identified (2022). The relative timing of studies relative to dominant VOC in the region where participants are recruited is important context for a complete picture of efficacy. Therefore, the efficacy and/or effectiveness of vaccines in the context of these variants is discussed where information is available.

Beyond the variability introduced by time and geography, efficacy within a trial and effectiveness in the real-world setting can also differ due to cohort differences. Patients participating in a clinical trial are likely to receive more medical oversight, resulting in better follow-up, adherence, and patient engagement ([13]). Additionally, the criteria for participant inclusion in a trial often bias trials towards selection of younger, healthier individuals ([14]). The ability of an RCT to accurately assess safety can be biased by the fact that a clinical trial might not reveal rare adverse events (AEs) that might become apparent on a larger scale ([14]). Therefore, while clinical trials are the gold standard for evaluating vaccines for COVID-19, the results of these trials must be considered in a broader context when real-world data is available.

While the relationship between a vaccine and a pathogen is not static, the data clearly demonstrates that a variety of efficacious vaccines have been developed against SARS-CoV-2. Here we discuss a selection of programs that use well-established vaccine biotechnologies. These programs have been undertaken worldwide, in complement to the more cutting-edge approaches developed and distributed in the United States (U.S.), the European Union (E.U.), the United Kingdom (U.K.), and Russia ([6]). In this review, we discuss vaccine development using two well-established technologies, whole-virus vaccination and subunit vaccination, and the role these technologies have played in the global response to the COVID-19 pandemic.

# 0.4 Whole-Virus Vaccines

Whole-virus vaccines have the longest history among vaccine development approaches. Variolation, which is widely considered the first vaccination strategy in human history, is one example ([15], [16]). Famously, variolation was employed against smallpox when healthy individuals were exposed to pus from an individual infected with what was believed to be either cowpox or horsepox ([15]–[18]). This approach worked by inducing a mild case of a disease. Therefore, while whole-virus vaccines confer adaptive immunity, they also raise safety concerns ([17], [19], [20]). As of 2005, most vaccines still used whole-virus platforms ([21]), and these technologies remain valuable tools in vaccine development today ([2]). Whole virus vaccine candidates have been developed for COVID-19 using both live attenuated viruses and inactivated whole viruses.

## 0.4.1 Live-Attenuated Virus Vaccines

Live-attenuated virus vaccines (LAV), also known as replication-competent vaccines, use a weakened, living version of a disease-causing virus or a version of a virus that is modified to induce an immune response ([5]). Whether variolation is the first example of a LAV being used to induce immunity is debated ([2](#), [19](#)). The first deliberate (albeit pathogen-naïve) attempt to develop an attenuated viral vaccine dates back to Louis Pasteur's efforts in 1885 to inoculate a child against rabies ([22](#)). The next intentional LAVs were developed against the yellow fever virus in 1935 and influenza in 1936 ([23](#)).

Early efforts in LAV development relied on either the identification of a related virus that was less virulent in humans (e.g., cowpox/horsepox or rotavirus vaccines) or the culturing of a virus *in vitro* ([2](#), [17](#)). Today, a virus can be attenuated by passaging it in a foreign host until, due to selection pressure, the virus loses its efficacy in the original host. Alternatively, selective gene deletion or codon deoptimization can be utilized to attenuate the virus ([5](#)), or foreign antigens can be integrated into an attenuated viral vector ([24](#)). LAVs tend to be restricted to viral replication in the tissues around the location of inoculation ([23](#)), and some can be administered intranasally ([5](#)).

Today, LAVs are used globally to prevent diseases caused by viruses such as measles, mumps, rubella, polio, influenza, varicella zoster, and the yellow fever virus ([25](#)). There were attempts to develop LAVs against both SARS-CoV-1 and MERS-CoV ([26](#)), but no vaccines were approved. It is generally recognized that LAVs induce an immune response similar to natural infection, and they are favored because they induce long-lasting and robust immunity that can protect from disease. This strong protective effect is induced in part by the immune response to the range of viral antigens available from LAV, which tend to be more immunogenic than those from non-replicating vaccines ([19](#), [26](#), [27](#)).

## 0.4.2 LAV Vaccines and COVID-19

To date, LAVs have not been widely deployed against SARS-CoV-2 and COVID-19. All the same, there are several COVID-19 LAV candidates in the early (preclinical/phase I) stages of investigation. These candidates utilize different approaches. Interestingly, several candidates (Meissa Vaccines' MV-014-212 and Codagenix's COVI-VAC, specifically) are administered intranasally, potentially improving accessibility by eliminating the need for sterile needles and reducing manufacturing costs, targeting conferring mucosal immunity, and reducing some issues related to vaccine hesitancy ([28](#), [29](#)). Additionally, although no phase III trial data is available for LAV vaccine candidates, some manufacturers have proactively sought to respond to the emergence of VOC. Therefore, the original approach to vaccination may still prove extremely advantageous in the high-tech landscape of COVID-19 vaccine development.

### 0.4.2.1    YF-S0

One candidate in the preclinical stage is YF-S0, a single-dose LAV developed at Belgium's Katholieke Universiteit Leuven that uses live-attenuated yellow fever 17D (YF17D) as a vector for a noncleavable prefusion conformation of the spike antigen of SARS-CoV-2 ([24](#)). YF-S0 induced a robust immune response in three animal models and prevented SARS-CoV-2 infection in macaques and hamsters ([24](#)). Additionally, the protective effect of YF-S0 against VOC has been investigated in hamsters ([30](#)). Even for a small number of hamsters that developed breakthrough infections after exposure to the index strain or the Alpha variant, viral loads were very low ([30](#)). However, much higher rates of breakthrough infection and higher viral loads were observed when the hamsters were exposed to the Beta variant ([30](#)). Reduced seroconversion and nAb titers were also observed against the Beta and Gamma variants ([30](#)). As a result, a modified version of YF-S0, called YF-S0*, was developed to include a modified spike protein intended to increase immunogenicity by including the full spectrum of amino acids found in the Gamma VOC as well as stabilizing the S protein's conformation ([30](#)). The updated vaccine was again tested in Syrian golden hamsters ([30](#)). No breakthrough infections were observed following vaccination with YF-S0* and exposure to the index strain and the Alpha, Beta, Gamma, and Delta variants ([30](#)). YF-S0* also reduced the infectious viral load in the lungs of several VOCs (Alpha, Beta, Gamma, and Delta) relative to a sham comparison ([30](#)), and the likelihood of the Delta variant spreading to unvaccinated co-housed hamsters was significantly reduced by YF-S0* ([30](#)). The updated vaccine was also associated with the increased production of nAbs against the Omicron variant compared to YF-S0 ([30](#)).

### 0.4.2.2    COVI-VAC

Other programs are developing codon deoptimized LAV candidates ([31](#)–[33](#)). This approach follows the synthetic attenuated virus engineering (SAVE) strategy to select codon substitutions that are suboptimal for the virus ([33](#), [34](#)). New York-based Codagenix and the Serum Institute of India reported a successful preclinical investigation ([33](#)) of an intranasally administered deoptimized SARS-CoV-2 LAV known as COVI-VAC, and COVI-VAC entered phase I human trials and dosed its first participants in January 2021 ([32](#), [35](#)). This vaccine is optimized through the removal of the furin cleavage site (see ([3](#)) for a discussion of this site's importance) and deoptimization of 283 codons ([36](#)). The results of the COVI-VAC phase I human trials are expected soon ([35](#)).

Other results suggest both potential benefits and risks to the COVI-VAC vaccine candidate. Preclinical results suggest that the vaccine candidate may confer some protection against VOC even though it was designed based on the index strain: a poster reported that Syrian golden hamsters who received COVI-VAC were significantly less likely to lose weight following viral challenge with the Beta VOC ([36](#)). On the other hand, some concerns have arisen about the possibility of spillover from LAV vaccines. A December 2022 study analyzed SARS-CoV-2 samples isolated from COVID-19 patients in India and identified two extremely similar sequences collected on June 30, 2020 that showed a high level of recombination relative to the dominant strains at the time ([37](#)). Comparing these samples to a database of SARS-CoV-2 sequences revealed they were most similar to the sequence used for COVI-VAC ([37](#)). Based on phylogenetic reconstruction, the authors argued that these SARS-CoV-2 isolates were most likely to have spilled over from COVI-VAC trials ([37](#)). If this was a case of spillover, the effect seems to have been limited, as these sequences were just two among over 1,600 analyzed. However, these concerns may be one consideration in the development of LAV vaccines for COVID-19.

### 0.4.2.3    Meissa Vaccines MV-014-212

Another company, Meissa Vaccines in Kansas, U.S., which also develops vaccines for respiratory syncytial virus (RSV), has developed an intranasal live-attenuated chimeric vaccine MV-014-212 ([38](#)). Chimeric vaccines integrate genomic content from multiple viruses to create a more stable LAV ([39](#)). To develop a SARS-CoV-2 vaccine candidate, Meissa Vaccines built on their prior work developing RSV vaccines ([38](#)). A live attenuated recombinant strain of RSV previously investigated as a vaccine candidate was modified to replace two surface glycoproteins with a chimeric protein containing components of the SARS-CoV-2 Spike protein as well as the RSV fusion (F) protein ([38](#)). Preclinical results describing the intranasal administration of MV-014-212 to African green monkeys and mice indicated that the vaccine candidate produced neutralizing antibodies (nAb) as well as a cellular immune response to SARS-CoV-2 challenge, including the Alpha, Beta, and Delta VOC ([38](#)). Enrollment for phase I human trials began in March 2021 and recruitment is ongoing ([32](#), [40](#)).

#### 0.4.2.4 Bacillus Calmette-Guerin Vaccines

Finally, Bacillus Calmette-Guerin (BCG) vaccines that use LAVs are being investigated for the prophylaxis of COVID-19 (see online Appendix ([41](#))). The purpose of the BCG vaccine is to prevent tuberculosis, but non-specific effects against other respiratory illnesses have suggested a possible benefit against COVID-19 ([42](#)). However, a multicenter trial that randomly assigned participants 60 years and older to vaccination with BCG (n = 1,008) or placebo (n = 1,006) found that BCG vaccination had no effect on the incidence of SARS-CoV-2 or other respiratory infections over the course of 12 months ([43](#)). Despite these findings, BCG vaccination was associated with a stronger cytokine (specifically, IL-6) response following *ex vivo* stimulation of peripheral blood mononuclear cells in patients with no known history of COVID-19 ([43](#)). Additionally, SARS-CoV-2-positive individuals who had received the BCG vaccine one year prior showed increased immunoglobulin (Ig) responses to the SARS-CoV-2 spike protein and receptor binding domain (RBD) relative to individuals who had received a placebo vaccine ([43](#)). Currently, investigations of BCG vaccines against COVID-19 are being sponsored by institutes in Australia in collaboration with the Bill and Melinda Gates Foundation ([44](#)) and by Texas A&M University in collaboration with numerous other U.S. institutions ([45](#)).

#### 0.4.2.5 Summary of LAV Vaccine Development

LAV vaccines for COVID-19 have not advanced as far in development as vaccines developed using other technologies. As of December 2022, COVI-VAC was the only LAV vaccine candidate in phase III clinical trials ([46](#)). As a result, safety data is not yet available for human studies of these vaccines. In general, though, safety concerns previously associated with LAV have been largely mitigated in the modern manufacturing process. Manufacturers use safe and reliable methods to produce large quantities of vaccines once they have undergone rigorous preclinical studies and clinical trials to evaluate their safety and efficacy. However, one remaining safety concern may be contributing to the relatively slow emergence of LAV candidates against COVID-19: they still may present risk to individuals who are immunocompromised ([47](#)), which is an even greater concern when dealing with a novel virus and disease. Additional data are needed to ascertain how this technology performs in the case of SARS-CoV-2 and whether rare cases of spillover have indeed occurred. Additionally, modifications to the design of individual vaccine candidates may make this protection more robust as SARS-CoV-2 evolves, as the limited data about LAV performance against VOC suggests. Despite the long and trusted history of LAV development, this vaccine strategy has not been favored against COVID-19, as other technologies have shown greater expediency and safety compared to the time-consuming nature of developing LAVs for a novel virus.

## 0.5  Inactivated Whole-Virus Vaccines

Inactivated whole-virus (IWV) vaccines are another well-established vaccine platform. This platform uses full virus particles generally produced via cell culture that have been rendered non-infectious by chemical (i.e., formaldehyde or β-propiolactone ([48])) or physical (i.e., heat or ultraviolet radiation) means. In general, these vaccines mimic the key properties of the virus that stimulate a robust immune response, but the risk of adverse reactions is reduced because the virus is inactivated and thus unable to replicate. Though these viral particles are inactivated, they retain the capacity to prime the immune system. The size of the viral particle makes it ideal for uptake by antigen-presenting cells, which leads to the stimulation of helper T-cells ([49]). Additionally, the array of epitopes on the surface of the virus increases antibody binding efficiency ([49]). The native conformation of the surface proteins, which is also important for eliciting an immune response, is preserved using these techniques ([50]). Membrane proteins, which support B-cell responses to surface proteins, are also induced by this method ([51]).

IWV vaccines have been a valuable tool in efforts to control many viruses. Some targets of IWV vaccines have included influenza viruses, poliovirus, and hepatitis A virus. Inactivated vaccines can generally be generated relatively quickly once the pathogenic virus has been isolated and can be passaged in cell culture ([26], [52]). During COVID-19, though the World Health Organization (WHO) has been slower to approve IWV vaccine candidates than those developed with nucleic acid-based technologies, IWV vaccine development was also fast. In China, the first emergency use authorization (EUA) was granted to an IWV vaccine in July 2020, with full approval following that December ([53], [54]). The fact that these vaccines have not received as much public attention (at least in Western media) as nucleic acid vaccines for SARS-CoV-2 may be due at least in part to the novelty of nucleic acid vaccine technologies ([55]), which are more modular and immunogenic ([6]).

Past applications to human coronaviruses (HCoV) have focused predominantly on SARS-CoV-1. Preclinical studies have demonstrated that IWV SARS-CoV-1 vaccine candidates elicited immune responses *in vivo*. These vaccines generated nAb titers at concentrations similar to those evoked by recombinant protein vaccines ([50], [56]). Studies in ferrets and non-human primates demonstrated that IWV vaccines can offer protection against infection due to nAb and SARS-CoV-1-specific T cell responses ([57]). However, several attempts to develop IWV vaccines against both SARS-CoV-1 and MERS-CoV were hindered by incidences of vaccine-associated disease enhancement (VADE) in preclinical studies ([58]). In one example of a study in macaques, an inactivated SARS-CoV-1 vaccine induced even more severe lung damage than the virus due to an enhanced immune reaction ([59]). Independent studies in mice also demonstrated evidence of lung immunopathology due to VADE in response to MERS-CoV IWV vaccination ([60], [61]). The exact mechanisms responsible for VADE remain elusive due to the specificity of the virus-host interactions involved, but VADE is the subject of investigation in preclinical SARS-CoV-2 vaccine studies to ensure the safety of any potential vaccines that may reach phase I trials and beyond ([58]).

## 0.5.1 Application to COVID-19

*Table 1: Inactivated whole-virus vaccines approved in at least one country ([62](#)) as of December 2, 2022 ([7](#)).*

| Vaccine | Company |
|---|---|
| Covaxin | Bharat Biotech |
| KoviVac | Chumakov Center |
| Turkovac | Health Institutes of Turkey |
| FAKHRAVAC (MIVAC) | Organization of Defensive Innovation and Research |
| QazVac | Research Institute for Biological Safety Problems (RIBSP) |
| KCONVAC | Shenzhen Kangtai Biological Products Co |
| COVIran Barekat | Shifa Pharmed Industrial Co |
| Covilo | Sinopharm (Beijing) |
| Inactivated (Vero Cells) | Sinopharm (Wuhan) |
| CoronaVac | Sinovac |
| VLA2001 | Valneva |

Several whole-virus vaccines have been developed against COVID-19 and are available in countries around the world (Table [1](#)). As of January 18, 2023, 10 vaccines developed with IWV technology are being distributed in 118 countries (Figure [1](#)). Evidence about the value of these vaccines to combat SARS-CoV-2 is available not only from clinical trials, but also from their roll-out following approval. Here, a major consideration has been that vaccines often lose efficacy as mutations accumulate in the epitopes of the circulating virus; IWV vaccines may be particularly affected in such cases ([20](#)). This loss of specificity over time is likely to be influenced by the evolution of the virus, and specifically by the rate of evolution in the region of the genome that codes for the antigenic spike protein. Here we review three vaccine development programs and their successes in a real-world setting.

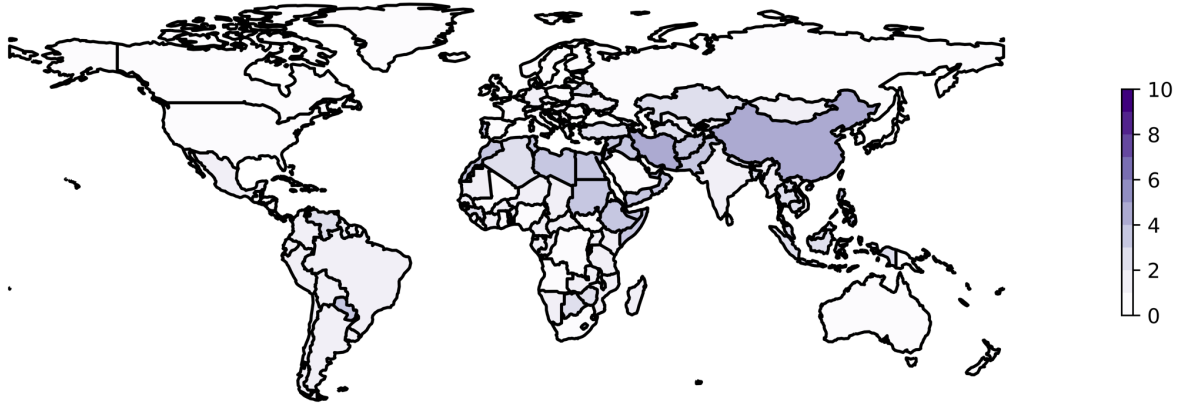

*Figure 1:* ***Worldwide availability of vaccines developed using inactivated whole viruses.*** *This figure reflects the number of vaccines based on whole inactivated virus technology that were available in each country as of January 18, 2023. These data are retrieved from Our World in Data ([63](#)) and plotted using geopandas ([64](#)). The color scale is based on the number of vaccines of this type included in the OWID dataset as a whole, not the maximum observed in a single country. See https://greenelab.github.io/covid19-review/ for the most recent version of this figure, which is updated daily.*

### 0.5.1.1 Sinovac's CoronaVac

One IWV vaccine, CoronaVac, was developed by Beijing-based biopharmaceutical company Sinovac. The developers inactivated a SARS-CoV-2 strain collected in China with β-propiolactone and propagated it using Vero cells ([26](#)). The vaccine is coupled with an aluminum adjuvant to increase immunogenicity ([26](#)). Administration follows a prime-boost regimen using a 0.5 ml dose containing 3 µg of inactivated SARS-CoV-2 virus per dose ([65](#)). In phase I and II clinical trials, CoronaVac elicited a strong immunogenic response in animal models and the development of nAbs in human participants ([66](#)–[68](#)). The phase I/II clinical trials were conducted in adults 18-59 years old ([68](#)) and adults over 60 years old ([66](#)) in China. Safety analysis of the CoronaVac vaccine during the phase II trial revealed that most adverse reactions were either mild (grade 1) or moderate (grade 2) in severity ([68](#)). In adults aged 18 to 59 years receiving a variety of dosage schedules, site injection pain was consistently the most common symptom reported ([68](#)). In older adults, the most common local and systemic reactions were pain at the injection site (9%) and fever (3%), respectively ([66](#)).

As of December 2022, a total of 17 CoronaVac trials had been registered in a variety of countries including the Philippines and Hong Kong ([69](#)). Two of the earliest phase III trials to produce results examined a two-dose regimen of CoronaVac following a 14-day prime boost regimen ([70](#), [71](#)). These trials were conducted in Turkey ([71](#)) and Chile ([70](#)) and enrolled participants over an identical period from September 2020 and January 2021. The Chilean trial, which reported interim results regarding safety and immunogenicity, identified specific IgG nAbs against the S1 RBD and a robust IFN-γ secreting T cell response was induced via immunization with CoronaVac ([70](#)). In the Turkish trial, VE was estimated to be 83.5% against symptomatic

COVID-19 ([71](#)). In the safety and immunogenicity study, minimal AEs were reported ([71](#)), and 18.9% of participants in the vaccine arm of the Turkish trial reported AEs compared to 16.9% of participants in the placebo group ([71](#)). However, 2% (n=7) of Turkish participants aged 18 to 59 reported severe AEs ([66](#)), causing the trial to be halted for investigation ([72](#)). The investigation determined that these events were unrelated to the vaccine ([66](#), [72](#)).

An additional phase III trial was conducted in Brazil between July and December 2020 following a randomized, multicenter, endpoint driven, double-blind, placebo-controlled design and enrolling nearly 10,000 healthcare workers ([73](#), [74](#)). The preprint reporting the results of this study ([74](#)) reports an efficacy of 50.7% against symptomatic COVID-19 and 100% against moderate to severe cases. A large percentage of participants, 77.1% in the vaccine group and 66.4% in the placebo group, reported AEs, including two deaths, but all of the serious AEs were determined not to be related to the vaccine ([74](#)). CoronaVac also appears to be suitable for use in immunocompromised patients such as those with autoimmune rheumatic diseases according to phase IV trials ([75](#)), and the vaccine was also well tolerated and induced humoral responses in phase I trials in children aged 3 to 17 years, which will now be examined in phase II and III clinical trials ([76](#)).

Estimates of CoronaVac's VE have varied across trials. The 50.7% VE reported from the Brazilian trial was contested by Turkish officials, as at the time the efficacy in the Turkish trial appeared to be 91.25% ([77](#), [78](#)). Ultimately, after multiple announcements, the efficacy debate was settled at over 50% ([77](#), [78](#)). Subsequently, the VE for the Turkish trial was finalized at 83.5% ([71](#)), and a prospective national cohort study in Chile reported an adjusted estimated effectiveness of 66% for the prevention of COVID-19 with an estimated 90% and 87% prevention of hospitalization and death, respectively ([79](#)).

Based on these results, CoronaVac was approved in China, and has now been distributed in 66 countries across Africa, Asia, Europe, North America, and South America, including Brazil, Cambodia, Chile, Colombia, Laos, Malaysia, Mexico, Turkey, Ukraine, and Uruguay ([8](#), [80](#)). As of August 2021, Sinovac had reportedly produced over a billion doses of CoronaVac ([80](#)). Outside of trials, rare cases of VADE have been reported in association with the CoronaVac vaccine ([81](#)). In one case study, two male patients both presented with COVID-19 pneumonia following vaccination with CoronaVac ([81](#)). This study identified the timeline of disease presentation, vaccination, and known COVID-19 exposure in the two patients and suggests that the inflammatory response induced by the vaccine could have caused an asymptomatic case of COVID-19 to present with symptoms ([81](#)). However, no causal relationship between CoronaVac and COVID-19 symptom onset was evaluated, and the reports are extremely rare.

The effectiveness has also been questioned based on real-world data, such as when concerns were raised about the vaccine's effectiveness following reports that over 350 doctors became ill with COVID-19 in Indonesia despite being immunized with CoronaVac ([82](#)). One possible explanation for such outbreaks was the evolution of the virus. Sera from individuals vaccinated with CoronaVac was found to show reduced neutralizing activity against the Alpha, Beta, and Delta VOC relative to the index strain ([83](#)). Similarly, a second study of 25 patients in Hong Kong in late 2021 evaluated serum neutralizing activity against the index strain and the Beta VOC, Delta VOC, and two Omicron isolates ([84](#)). They reported that all individuals were seropositive for nAbs against the index strain, 68% against the Delta variant, and 0% against the Beta VOC and Omicron isolates ([84](#)). The Beta variant appears to be more resistant to nAbs in sera from individuals immunized with CoronaVac than the Alpha variant or wildtype virus, indicating that emerging variants may be of concern ([85](#)). Finally, a fourth study examined sera from 180 Thai healthcare workers vaccinated with CoronaVac and reported that neutralizing activity was significantly reduced against the Alpha, Delta, and Beta variants relative to the index strain ([86](#)). Together, these results suggest that viral evolution is likely to pose a significant challenge to immunity acquired from the CoronaVac vaccine.

Therefore, studies have also evaluated whether booster doses would provide additional protection to individuals vaccinated with CoronaVac. This strategy is supported by the fact that the antibody response elicited by CoronaVac has been found to wane following the second dose, though it was still detected six months out ([87](#)). A phase I/II clinical trial of CoronaVac in an elderly cohort (adults 60 years and older) in China determined that by 6 to 8 months following the second dose, nAb titers were detected below the seropositive cutoff ([88](#)). Data from two phase II trials indicated that nAb response had declined 6 months after the second dose of the primary series, but a booster dose of CoronaVac administered 8 months after the second dose markedly increased geometric mean titers of SARS-CoV-2 nAbs ([89](#)). However, the reduction of nAbs was ameliorated by a booster dose administered 8 months after the second CoronaVac dose ([89](#)). Furthermore, Chinese ([90](#)) and Chilean ([91](#)) researchers have opted to investigate options to administer different vaccines (e.g., an mRNA vaccine dose) as a booster dose to individuals who have already received two doses of the IWV vaccine CoronaVac. Another study determined that using a viral-vectored vaccine (CanSino's Convidecia) or an mRNA vaccine (Pfizer/BioNTech's BNT162b2) instead of CoronaVac in a prime-boost vaccination regimen could induce a more robust immune response ([92](#), [93](#)). The WHO now suggests that a booster dose, either homologous or heterologous, can be considered 4 to 6 months after the primary series, especially for high-risk groups ([94](#)).

### 0.5.1.2    Sinopharm's Covilo

Two additional IWV vaccine candidates were developed following a similar approach by the state-owned China National Pharmaceutical Group Co., Ltd., more commonly known as Sinopharm CNBG. One, known as BBIBP-CorV or Covilo, was developed in Beijing using the HB02 variant of SARS-CoV-2. The other was developed at Sinopharm CNBG's Wuhan Institute using the WIV04 variant of SARS-CoV-2 ([95](#)). The viruses were purified, propagated using Vero cells, isolated, and inactivated using β-propiolactone ([95](#), [96](#)). Both vaccines are adjuvanted with aluminum hydroxide ([95](#), [96](#)). Here, we focus on Covilo.

Preclinical studies indicated that Covilo induced sufficient nAb titers in mice, and a prime-boost immunization scheme of 2 μg/dose was sufficient to protect rhesus macaques from disease ([96](#)). In phase II trials, the Covilo vaccine appeared well-tolerated, with 23% of participants in the vaccine condition (482 total participants, 3:1, vaccine:placebo) reporting at least one adverse reaction characterized as mild to moderate ([97](#)). No evidence of VADE was detected using this vaccine in phase II data ([98](#)). In phase III trials conducted between July and December 2020, Covilo achieved an efficacy of 72.8% and was well tolerated ([99](#)). However, questions were raised about efficacy when Sinopharm affiliates in the UAE in early December 2020 claimed the vaccine had 86% efficacy, which was later at odds with a Sinopharm Beijing affiliate that stated that Covilo had a 79.34% efficacy later that same month ([100](#)).

Studies have also investigated expected differences in real-world effectiveness of Covilo given the continuing evolution of SARS-CoV-2. The antibody response elicited by Covilo was found to wane, but still to be detectable, by six months following the second dose (87). One study showed that the Alpha variant exhibited very little resistance to neutralization by sera of those immunized with Covilo, but the Beta variant was more resistant to neutralization by almost a factor of 3 (85). Another study examined sera from 282 participants and used a surrogate neutralizing assay, a test that generally correlates with nAbs, to determine that Covilo appears to induce nAbs against the binding of the RBD of wild type SARS-CoV-2 and the Alpha, Beta, and Delta variants to ACE2 (101). Notably, a preprint reported that antisera (i.e., the antibody-containing component of the sera) from 12 people immunized with Covilo exhibited nAb capacity against the Beta variant (B.1.351), wild type SARS-CoV-2 (NB02), and one of the original variants of SARS-CoV-2 (D614G) (102). As with many other vaccines, booster doses are being evaluated to mitigate some of the issues arising from viral evolution. A study of healthcare workers in China has since indicated that a booster shot of Covilo elevates B cell and T cell responses and increases nAb titers (103). In May 2021, the UAE announced it would consider booster shots for all citizens who had been immunized with Covilo, which was shortly followed by a similar announcement in Bahrain, and by August 29, 2021, the UAE mandated booster shots for all residents who had received Covilo (80).

### 0.5.1.3    Bharat Biotech's Covaxin

Another IWV vaccine candidate was developed by Bharat Biotech International Ltd., which is the biggest producer of vaccines globally, in collaboration with the Indian Council of Medical Research (ICMR) National Institute of Virology (NIV). This candidate is known as Covaxin or BBV152. Preclinical studies of Covaxin in hamsters (104) and macaques (105) indicated that the vaccine induced protective responses deemed sufficient to move forward to human trials. Phase I (July 2020) and phase I/II (September to October 2020) studies indicated that Covaxin adjuvanted with alum and a Toll-like receptor 7/8 (TLR7/8) agonist was safe and immunogenic (106, 107). These two studies demonstrated that the vaccine induced significant humoral and cell-mediated responses, as assessed by measuring binding (106) and neutralizing (106, 107) antibodies, cytokines (106, 107), CD3$^+$, CD4$^+$, and CD8$^+$ T-cells (106), with some formulations also eliciting Th1-skewed memory T-cell responses (107). Only mild to moderate side-effects were reported upon immunization (106, 107), and in phase II trials, the Covilo vaccine appeared well-tolerated (97).

In India, the Covaxin vaccine received emergency authorization on January 3, 2021, but the phase III data was not released until March 3, 2021, and even then it was communicated via press release (108). This press release reported 80.6% efficacy in 25,800 participants (108, 109), spurring Zimbabwe to follow suit and authorize the use of Covaxin (110). A detailed preprint describing the double-blind, randomized, controlled phase III trial that enrolled between November 2020 and January 2021 became available in July 2021 (111), and the results collected as of May 17, 2021 were published in December 2021 (112). Based on a final enrollment of 25,798 people (~1:1 vaccine:placebo), overall VE against symptomatic COVID-19 was estimated at 77.8% and against severe disease and asymptomatic infection was reported as 93.4% and 63.6%, respectively (112). The vaccine was also reported to be well tolerated, with fewer severe events occurring in the Covaxin group (0.3%) than in the placebo group

(0.5%) ([112](#)). One case of a serious AE potentially related to the vaccine, immune thrombocytopenic purpura, was reported, although this patient was seropositive for SARS-CoV-2 at the baseline observation point ([112](#)). As of January 19, 2023, Covaxin was approved for emergency use in 31 countries across Africa, Asia, Europe, and South America, including Guyana, India, Iran, Zimbabwe, Nepal, Mauritius, Mexico, Nepal, Paraguay, and the Philippines ([113](#)).

Like with all vaccines, the continued evolution of SARS-CoV-2 poses a challenge to the effectiveness of Covaxin. In this case, the phase III clinical trial did evaluate the efficacy of Covaxin in response to variants circulating in mid-to-late 2020 ([112](#)). In agreement with previous studies demonstrating sera from individuals vaccinated with Covaxin efficiently neutralized the Alpha variant (B.1.1.7) and the Delta variant (B.1.617.2) ([114](#)–[116](#)), the phase III trial reported a 65.2% efficacy against the Delta variant (B.1.617.2) ([112](#)). Another study reported that sera from individuals immunized with Covaxin produced effective nAbs against the Delta variant and the so-called Delta plus variant (AY.1) ([117](#)). Indeed, sera obtained from Covaxin boosted individuals (n=13) ([118](#)) and those who were vaccinated with Covaxin but recovered from a breakthrough infection (n=31) also neutralized the Omicron variant ([119](#)). Therefore, the data suggest that the vaccine does continue to confer protection to VOC.

The authorization of Covaxin has also offered opportunities to monitor how well the clinical trial results translate into a real-world setting. Additionally, an effort to monitor AEs and COVID-19 cases following vaccine roll-out reported that most side effects were mild and that cases were rare, even though this data would seem to have been collected during the severe wave of COVID-19 brought on by the Delta VOC in India in early 2021; at the same time, the sample sizes were extremely small ([120](#)). Similarly, larger studies of adults (June to September 2021) ([121](#)) and adolescents (beginning in January 2022) ([122](#)) who received the vaccine outside of a trial setting reported that safety was similar across age groups, with no severe AEs reported in adults and with no serious AEs reported in adolescents, although 0.9% (6 individuals) reported severe AEs. However, a much lower effectiveness (22-29%) was estimated in a real-world setting during an analysis of cases in healthcare workers from April to May 2021 ([123](#)). All the same, monitoring of hospitalized COVID-19 patients between April and June 2021 indicated that the vaccines were highly effective against preventing severe illness ([124](#)).

It is not yet clear what level of protection Covaxin offers beyond 6 to 8 months post the second vaccine; consequently, the potential requirement of a booster immunization is being explored ([125](#)). Furthermore, Bharat Biotech is considering other vaccine regimens such as providing one initial immunization with Covaxin followed by two immunizations with its intranasal vaccine (BBV154) ([126](#)). U.S.-based Ocugen Inc., a co-development partner of Bharat Biotech, is leading the application for an Emergency Use Authorization (EUA) for Covaxin intended for the U.S. market. It has been reported that Bharat Biotech will soon release its phase II and III pediatric trial results ([127](#)).

However, the WHO approval of the Covaxin has been delayed ([128](#)), and in April 2022, the WHO suspended procurement of Covaxin due to concerns about deviation from good manufacturing practice in their production facilities ([129](#), [130](#)). All the same, no safety issues had been reported in association with this vaccine, and the suspension was unlikely to affect distribution given that Bharat Biotech had not been supplying doses through this mechanism ([131]). Clinical trials had recommenced in the United States as of May 2022 ([131]).

## 0.5.2 Summary of IWV Vaccine Development

In the past, problems that arose during the manufacturing of IWV vaccines could present safety issues, but oversight of the manufacturing process has helped to improve IWV vaccine safety ([132]). Nevertheless, the departure from norms necessitated by the COVID-19 crisis raised concerns about whether oversight would occur at pre-pandemic standards ([132]). In general, the IWV COVID-19 vaccines have reported very few issues with safety. Additionally, safety audits have proactively identified concerns, as demonstrated with the WHO's suspension of Covaxin.

More concern has arisen around the issue of effectiveness due to the reduced neutralizing activity of IWV vaccines against VOC relative to the index strain. In several cases, estimates of VE have varied widely across different trials of a single vaccine. Such issues are likely to be exacerbated by spatiotemporal differences in viral evolution, though in the case of the very high estimate generated by the Turkish trial of CoronaVac ([71]), the design of the study may have inflated the VE estimate ([133]). Regardless, the authors of the original trial argued that all of the trials suggest a very high efficacy against severe disease ([134]), as is the case for all of the IWV vaccines discussed here. In addition to issues related to the evolution of SARS-CoV-2, it is important to consider the duration of immunity over time. With IWV vaccines, heterologous vaccine boosters are being considered in many cases. Today, the WHO has developed recommendations for booster immunization for several whole-virus vaccines. In some cases (Valneva-VLA2001 ([135]), Covaxin ([136]), Covilo ([137]), Sinopharm-WIBP Inactivated (Vero Cell) ([138])), boosters are recommended only for high-risk and/or high-priority groups (e.g., the immunocompromised and medical professionals, respectively), while for Sinovac's CoronaVac ([94]), they are recommended more broadly. Studies are also investigating the effects of booster doses in other vaccines ([139]–[141]), though some are being investigated or deployed primarily as heterologous boosters in populations vaccinated with a different primary series ([140]).

As new vaccines are approved by the WHO, more time elapses since many received the primary series, and new variants emerge, booster recommendations are likely to increase. Therefore, IWV vaccines have played an important role in vaccine access during the initial phase of vaccination against COVID-19, but many IWV vaccinees may receive booster doses developed with emergent vaccine technologies like DNA and mRNA. In head-to-head comparisons, these types of vaccines were typically found to outperform IWV vaccines (e.g., ([84], [86], [144]). At the same time, IWV vaccines are among the easiest to store and transport due to requiring refrigeration only at 2 to 8°C and remaining stable for years at a time ([99]). Therefore, these vaccines are likely to continue to play an important role in vaccine equity and accessibility.

# 0.6 Subunit Vaccines

Efforts to overcome the limitations of live-virus vaccines led to the development of approaches first to inactivate viruses (circa 1900), leading to IWV vaccines, and then to purifying proteins from viruses cultured in eggs (circa 1920) ([2], [145]). The purification of proteins then set the stage for the development of subunit vaccines based on the principle that the immune system can be stimulated by introducing one or more proteins or peptides isolated from the virus. Today, such approaches often use antigens isolated from the surface of the viral particle that are key targets of the immune system (protein subunit vaccines). Advances in biological engineering have also facilitated the development of approaches like viral-like particle (VLP) vaccines using nanotechnology ([146](#)). VLPs share the conformation of a virus's capsid, thereby acting as an antigen, but lack the replication machinery ([147](#)).

Unlike whole-virus vaccines, which introduce the whole virus, subunit vaccines stimulate the immune system by introducing one or more proteins or peptides of the virus that have been isolated. The main advantage of this platform is that subunit vaccines are considered very safe, as the antigen alone cannot cause an infection ([148](#)). Both protein subunit and VLP vaccines thus mimic the principle of whole virus vaccines but lack the genetic material required for replication, removing the risk of infection ([149](#)). Protein subunit vaccines can stimulate antibodies and CD4$^+$ T-cell responses ([147](#), [150](#)).

The subunit approach is also favored for its consistency in production. The components can be designed for a highly targeted immune response to a specific pathogen using synthetic immunogenic particles, allowing the vaccine to be engineered to avoid allergen and reactogenic sequences ([151](#)). One limitation is that, in the case of protein subunit vaccines, adjuvants are usually required to boost the immune response ([152](#)) (see online Appendix ([41](#))). Adjuvants, which are compounds that elicit an immunogenic effect, include alum (aluminum hydroxide), squalene- or saponin-based adjuvants, and Freund's incomplete/complete adjuvants, although the latter is avoided in human and veterinary medicine due to high toxicity ([151](#), [153](#), [154](#)).

Protein subunit vaccine development efforts for both SARS-CoV-1 and MERS-CoV explored a variety of immunogens as potential targets. The search for a potential SARS-CoV-1 vaccine included the development of vaccines based on the full-length or trimeric S protein ([155](#)–[157](#)), those focused on the RBD protein only ([158](#)–[161](#)) or non-RBD S protein fragments ([156](#), [162](#)), and those targeting the N and M proteins ([163](#)–[165](#)). These efforts have been thoroughly reviewed elsewhere ([166](#)). There have been examples of successful preclinical research including candidate RBD219N-1, a 218-amino-acid residue of the SARS-CoV-1 RBD that, when adjuvanted to aluminum hydroxide, was capable of eliciting a high antibody response of both nAbs and RBD-specific monoclonal antibodies in both pseudovirus and live virus infections of immunized mice ([167](#)).

Similarly to the SARS-CoV-1 vaccine candidates, the MERS-CoV protein subunit vaccine candidates generally target the RBD ([158](#), [166](#), [168](#)–[171](#)), with some targeting the full length S protein ([172](#)), non-RBD protein fragments such as the SP3 peptide ([173](#)), and the recombinant N-terminal domain (rNTD) ([174](#)). Other strategies investigating the potential use of the full length S DNA have also been investigated in mice and rhesus macaques, which elicited immune responses ([175](#)), but these responses were not as effective as the combination of S DNA and the S1 subunit protein together ([175](#), [176](#)). No protein subunit vaccine for MERS-CoV has progressed beyond preclinical research to date. VLPs have been investigated for development of vaccines against MERS and SARS ([177](#), [178](#)) including testing in animal models ([179](#), [180](#)), but once again, only preclinical data against HCoV has been collected ([181](#)). However, protein subunit vaccines do play a role in public health and have contributed to vaccination against hepatitis B ([182](#)) and pertussis ([183](#), [184](#)) since the 1980s and human papillomavirus since 2006 ([185](#)). They are likely to continue to contribute to public health for the foreseeable future due to ongoing research in vaccines against influenza, SARS-CoV-2, Epstein-Barr virus, dengue virus, and human papillomavirus among others ([186](#)–[188](#)).

## 0.6.1 Application to COVID-19

*Table 2: Subunit vaccines approved for use in at least one country ([62](#)) as of December 2, 2022 ([7](#)).*

| Vaccine | Company | Platform |
| --- | --- | --- |
| Zifivax | Anhui Zhifei Longcom | protein subunit |
| Noora vaccine | Bagheiat-allah University of Medical Sciences | protein subunit |
| Corbevax | Biological E Limited | protein subunit |
| Abdala | Center for Genetic Engineering and Biotechnology (CIGB) | protein subunit |
| Soberana 02 | Instituto Finlay de Vacunas Cuba | protein subunit |
| Soberana Plus | Instituto Finlay de Vacunas Cuba | protein subunit |
| V-01 | Livzon Mabpharm Inc | protein subunit |
| Covifenz | Medicago | VLP |
| MVC-COV1901 | Medigen | protein subunit |
| Recombinant SARS-CoV-2 Vaccine (CHO Cell) | National Vaccine and Serum Institute | protein subunit |
| Nuvaxovid | Novavax | protein subunit |
| IndoVac | PT Bio Farma | protein subunit |
| Razi Cov Pars | Razi Vaccine and Serum Research Institute | protein subunit |
| VidPrevtyn Beta | Sanofi/GSK | protein subunit |
| COVOVAX (Novavax formulation) | Serum Institute of India | protein subunit |
| SKYCovione | SK Bioscience Co Ltd | protein subunit |
| TAK-019 (Novavax formulation) | Takeda | protein subunit |

| Vaccine | Company | Platform |
|---------|---------|----------|
| SpikoGen | Vaxine/CinnaGen Co. | protein subunit |
| Aurora-CoV | Vector State Research Center of Virology and Biotechnology | protein subunit |
| EpiVacCorona | Vector State Research Center of Virology and Biotechnology | protein subunit |

The development of subunit vaccines against SARS-CoV-2 is a remarkable achievement given the short period of time since the emergence of SARS-CoV-2 in late 2019, particularly considering these types of vaccines have not played a major role in previous pandemics compared to LAV and IWV vaccines. More than 20 protein subunit vaccines from companies such as Sanofi/GlaxoSmithKline, Nanogen, and the Serum Institute of India have entered clinical trials for COVID-19 since the beginning of the pandemic ([187]), 20 have been approved, and at least 9 are being administered worldwide ([7], [8]) (Table [2]). As of January 18, 2023, protein subunit vaccines are being distributed in at least 42 countries (Figure [2]).

Number of Protein Subunit Vaccines Available Worldwide

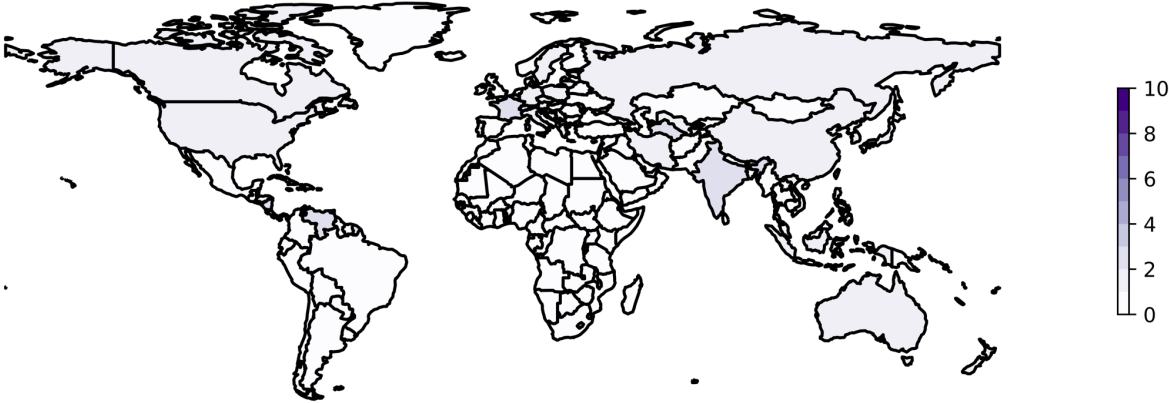

Figure 2: **Worldwide availability of vaccines developed using protein subunit.** *This figure reflects the number of vaccines based on protein subunit technology that were available in each country as of January 18, 2023. These data are retrieved from Our World in Data ([8], [63]) and plotted using geopandas ([64]). The color scale is based on the number of vaccines of this type included in the OWID dataset as a whole, not the maximum observed in a single country. See https://greenelab.github.io/covid19-review/ for the most recent version of this figure, which is updated daily.*

VLP vaccines have not progressed as rapidly. Programs seeking to develop VLP vaccines have used either the full-length S protein or the RBD of the S protein specifically as an antigen, although some use several different SARS-CoV-2 proteins ([148]). As of January 18, 2023, only one VLP was available in one country (Canada) ([8]).

Number of VLP Vaccines Available Worldwide

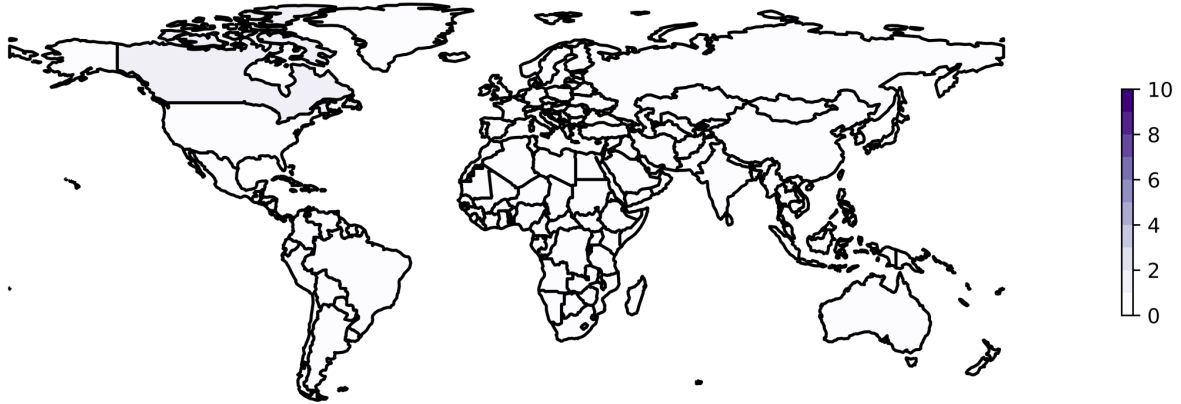

*Figure 3:* **Worldwide availability of vaccines developed with VLPs.** *This figure reflects the number of vaccines based on VLP technology that were available in each country as of January 18, 2023. These data are retrieved from Our World in Data ([63](#)) and plotted using geopandas ([64](#)). The color scale is based on the number of vaccines of this type included in the OWID dataset as a whole, not the maximum observed in a single country. See https://greenelab.github.io/covid19-review/ for the most recent version of this figure, which is updated daily.*

## 0.6.2 Novavax's Nuvaxovid

NVX-CoV2373, also known as Nuvaxovid or Covovax ([189](#)), is a protein subunit vaccine for SARS-CoV-2 produced by U.S. company Novavax and partners. Nuvaxovid is a protein nanoparticle vaccine constructed from a mutated prefusion SARS-CoV-2 spike protein in combination with a specialized saponin-based adjuvant to elicit an immune response against SARS-CoV-2 ([190](#)). The spike protein is recombinantly expressed in Sf9 insect cells ([191](#)), which have previously been used for several other FDA-approved protein therapeutics ([192](#)), and contains mutations in the furin cleavage site (682-RRAR-685 to 682-QQAQ-685) along with two proline substitutions (K986P and V987P) that stabilize the protein ([193](#)), including improving thermostability ([191](#)).

In preclinical mouse models, Nuvaxovid elicited high anti-spike IgG titers 21 to 28 days post-vaccination that could neutralize the SARS-CoV-2 virus and protect the animals against viral challenge, with particularly strong effects when administered with the proprietary adjuvant Matrix-M™ ([191](#)). In a phase I/II trial, a two-dose regimen of Nuvaxovid was found to induce anti-spike IgG levels and nAb titers exceeding those observed in convalescent plasma donated by symptomatic patients ([190](#)). In line with the preclinical studies, the use of Matrix-M adjuvant further increased anti-spike immunoglobulin levels and induced a Th1 response.

In a phase III randomized, observer-blinded, placebo-controlled clinical trial in the U.K., 14,039 participants received two 5-µg doses of Nuvaxovid or placebo administered 21 days apart in a 1:1 ratio from late September to late November 2020 ([194](#)). The primary endpoint of the trial was the occurrence or absence of PCR-confirmed, symptomatic mild, moderate or severe

COVID-19 from 7 days after the second dose onward ([194](#)). The VE was reported to be 89.7%, with a total of 10 patients developing COVID-19 in the vaccine group versus 96 in the placebo group ([194](#)). No hospitalizations or deaths were reported in the vaccine group ([194](#)). An additional phase III randomized, observer-blinded, placebo-controlled trial was conducted in the U.S. and Mexico, enrolling 29,949 participants and administering at least 1 vaccine in a 2:1 ratio from late December 2020 to late February 2021 ([195](#)). This trial ([195](#)) used the same primary endpoints as the initial phase III trial conducted in the U.K. ([194](#)). A vaccine efficacy of 90.4% was reported based on 77 cases total, 63 of which occurred in the placebo group ([195](#)). All moderate to severe cases of COVID-19 occurred in the placebo group ([195](#)). Hospitalization and death were not evaluated as individual secondary endpoints, but were instead included in the definition of severe COVID-19; all-cause mortality was comparable between the placebo and treatment conditions ([195](#)).

In both trials, the vaccine was found to be well-tolerated ([194](#), [196](#)). Analysis of 2,310 participants in the U.K. trial revealed that solicited AEs were much more common in the vaccine group than the placebo group across both doses, but the rate of unsolicited events was, while still higher in the vaccine group, much more similar ([194](#)). A small number of severe AEs were reported by vaccine recipients, including one case of myocarditis; however, the myocarditis was determined to be viral myocarditis ([194](#)). Common AEs were generally considered mild, with low incidences of headache, muscle pain, and fatigue ([195](#)). In the trial conducted in the U.S. and Mexico, once again, the most common symptoms included headache, fatigue, and pain, as well as malaise ([195](#)). Here, severe AEs were balanced across the vaccine and placebo groups ([195](#)). Thus, both trials suggested that the Nuvaxovid vaccine is safe and effective against COVID-19.

However, Novavax experienced significant challenges in preparing Nuvaxovid for distribution. Prior to the pandemic's onset, Novavax had sold their manufacturing facilities and reduced their staff dramatically ([197](#)). As a result, once they began producing Nuvaxovid, they struggled to establish a stable relationship with contractors who could produce the vaccine ([198](#)), especially given the challenge of producing vaccines at scale ([199](#)). Additionally, Novavax was not able to meet the purity standards laid out by the FDA ([200](#)). Eventually, the manufacturing issues were resolved ([201](#)), and Nuvaxovid has since been authorized by the WHO ([202](#)) and by political entities, including the United Kingdom ([203](#)), the E.U. ([204](#)), and the U.S. ([205](#)). These delays obstructed some of the goals of the vaccine development program, which was undertaken with significant investment from the U.S. government through Operation Warp Speed ([200](#)). Novavax was supposed to provide over a billion doses of Nuvaxovid to countries around the world through the COVID-19 Vaccines Global Access (COVAX) Facility ([200](#)). However, following the delays, Gavi (which oversees COVAX) terminated the agreement, leading to ongoing legal disagreements between the nonprofit and Novavax as of late 2022 ([201](#), [206](#)).

As with other vaccines, the question of how well Nuvaxovid continues to provide protection as SARS-CoV-2 evolves has been raised. *Post hoc* analysis in the phase III trial indicated a VE of 86.3% against the Alpha variant (identified based on the presence of the 69–70del polymorphism) and 96.4% against viral specimens lacking the 69-70del polymorphism ([194](#)). In the second phase III trial ([196](#)), whole-genome sequencing was obtained from 61 of the 77 observed cases, and 79% of infections were identified as a VOC or variant of interest (VOI) known at the time of the study. Vaccine efficacy against cases caused by VOC, among which the Alpha variant was predominant (88.6%), was reported to be 92.6% ([196](#)). In late 2020, an analysis of efficacy in South African adults revealed an overall efficacy of 60.1% and a slightly lower efficacy of 50.1% against the Beta variant (B.1.351) in particular ([207](#)).

The company has also initiated the development of new constructs to select candidates that can be used as a booster against new strains and plans to initiate clinical trials for these new constructs in the second quarter of 2021. An analysis of a booster dose of Nuvaxovid administered six months after the primary series revealed a significant increase in neutralizing activity against VOC including Delta and Omicron ([208](#)). This trial was conducted at 18 sites across the United States and Australia ([209](#)). Novavax has also initiated booster trials in the U.K. ([80](#)). Boosters may be especially important given that Omicron and related variants, in particular, may be associated with significantly reduced efficacy of Nuvaxovid ([210](#)).

Given the apparent need for boosters, interest has also emerged in whether booster doses of Nuvaxovid can be safely administered along with annual flu vaccines. In a subgroup of approximately 400 patients enrolled from the U.K. phase III trial who received either Nuvaxovid or a placebo at a ratio of 1:1, a concomitant dose of adjuvanted seasonal influenza vaccines (either a trivalent vaccine or a quadrivalent vaccine) was administered ([211](#)). This study demonstrated that the vaccines could be safely administered together ([211](#)). While no change to the immune response was noted for the influenza vaccine, a notable reduction of the antibody response elicited by Nuvaxovid was reported, but efficacy was still high at 87.5% ([211](#)). Novavax has since started phase I/II trials to investigate the administration of its own influenza vaccine, NanoFlu, concomitantly with Nuvaxovid ([212](#)). The combination appeared to be safe and effective in preclinical studies ([213](#)).

### 0.6.3 The Cuban Center for Genetic Engineering and Biotechnology's Abdala Vaccine

Another notable protein subunit vaccine development program came out of Cuba. Concerned about their ability to access vaccines, especially given the U.S.'s embargo ([214](#)), health officials in this developing country made the decision in March 2020 to undertake vaccine development on their own ([215](#)). Today, three Cuban protein subunit vaccines have been approved for use: Abdala, which was developed at the Cuban Genetic Engineering and Biotechnology Center and SOBERANA 02 and SOBERANA Plus, which were developed at Cuba's Finlay Vaccine Institute (Instituto Finlay de Vacunas Cuba) ([215](#)). Here, we focus on the development of the Abdala vaccine, but SOBERANA 01/02/Plus vaccine development program has also achieved great success and reported VEs of over 90% in the three-dose regimen ([216](#), [217](#)).

Abdala, also known as CIGB-66, was developed using yeast as a low-cost alternative to mammalian cell expression systems (e.g., human embryonic kidney cells) to cultivate the recombinant proteins that form the basis of this protein subunit vaccine ([218](#)). A sequence corresponding to the RBD of the Spike protein in the index strain of SARS-CoV-2 was codon optimized for expression in yeast, and the RBD proteins were then purified and used to inoculate mice, rats, and African green monkeys ([218](#)). In addition to the proteins, the vaccine candidate included an adjuvant, aluminum hydroxide gel ([218](#)). Comparing the immunogenicity of the yeast-cultivated proteins to those cultivated in human embryonic kidney cells revealed no significant difference in the immune response ([218](#)).

Based on promising results in laboratory animal testing, Abdala moved to phase I/II trials in human subjects ages 19 to 80, recruiting participants between December 2020 and February 2021 ([219](#)). The three-dose vaccine elicited no serious AEs across either phase I or II, and the vaccine was found to produce a strong immune response ([219](#)). In March 2021, phase III trials began ([220](#)), and by June, officials were reporting the VE to be 92.28% ([221](#), [222](#)). This high efficacy estimate, along with the short timeline of data collection, initially elicited skepticism, especially given that the data were not made public ([223](#)). However, the trials were designed to enroll a large number of participants and were carried out during a wave of infections due to the arrival of variants carrying the D614G mutation in Cuba, which would be expected to allow an expedited timeline for interim analysis ([223](#)). Based on the reported results, Abdala gained emergency use authorization in Cuba in July 2021 ([224](#)), and by December 2021, cases in Cuba had dropped dramatically ([225](#)). The results of the phase III trial were posted to *medRxiv* in September 2022, describing the results of a randomized, placebo-controlled, multicenter, double-blind investigation of the Abdala vaccine candidate in 48,000 participants between March 22 and April 3, 2021 ([226](#)). The final results evaluated 42 symptomatic cases of COVID-19 among participants in the placebo condition compared to only 11 cases among participants who received the vaccine, yielding the reported VE of 92.28% ([226](#)). In terms of secondary endpoints, the VE was 91.96% against mild/moderate COVID-19, 94.46% against severe COVID-19, and 100% against critical illness and death ([226](#)). The vaccine was also found to be very safe, with the overall incidence of AEs only 2.5% in vaccine recipients compared to 1.9% in the placebo recipients ([226](#)). Therefore, the phase III trial suggests that this vaccine is highly effective and safe.

Evidence from the deployment of the vaccine also suggests it is highly effective. A retrospective cohort study conducted between May and August 2021 evaluated public health data from over a million people in the city of Havana and found that the real-world effectiveness of the vaccine met or exceeded estimates of VE during the trial, with 98.2% effectiveness against severe disease and 98.7% effectiveness against death observed in fully vaccinated subjects ([227](#)). Notably, Cuba has vaccinated a high percentage of its population, with 87% of the population vaccinated by January 2022 and 90.3% by the end of December 2022 ([228](#), [229](#)). Therefore, one consideration in interpreting retrospective cohort studies is that the vaccination rate in Cuba is so high that the two cohorts might not be directly comparable. All the same, the fact that the efficacy and effectiveness of the Abdala vaccine have both been estimated to be over 90% against severe illness suggests that this vaccine is highly effective for mitigating the risk of COVID-19. As of December 2022, the vaccine had been authorized for distribution in five additional countries, including Mexico and Vietnam, although its evaluation for WHO approval was ongoing ([230](#), [231](#)).

However, limited data is available about the Abdala's vaccine's robustness to evolutionary changes in SARS-CoV-2. An *in silico* analysis identified several potential changes in the epitopes of the Omicron VOC relative to the sequence used in the development of Abdala ([232](#)). Instead, Cuban health officials have prioritized boosters. A representative of the Cuban state business group reportedly stated that immunity remains high at six months after the primary course but that some people may be prone to infection ([233](#)), suggesting waning immunity. The

Cuban government authorized boosters in January 2022 in an effort to mitigate the effects of the Omicron variant ([233]–[235]). Additional support for the efficacy of Abdala and other Cuban vaccines comes from the fact that Cuba's COVID-19 death rate has virtually flatlined since fall 2021, with less than 250 deaths reported during the entire year of 2022 in a population of 11.3 million ([236]). Therefore, in addition to developing a vaccine with an estimated VE paralleling that of vaccines developed using cutting-edge nucleic acid technologies ([6]), Cuba's vaccine roll out has also been much more successful than in nearly all similarly sized countries. This remarkable vaccine program underscores the continued importance of established, cost-effective vaccine development strategies ([234]) that make it possible for countries that have not traditionally been a leader in biotechnological innovation but have developed a solid vaccine production sector ([237]) to develop and produce vaccines that will serve their own population's needs. Additionally, Cuba's vaccines are uniquely accessible to many countries around the world ([234]).

## 0.6.4 Medicago's Covifenz

The leading example of a VLP approach applied to COVID-19 comes from Covifenz, developed by Canadian company Medicago ([238]). This vaccine was developed using plant-based VLP technology ([239]) that the company had been investigating in order to develop a high-throughput quadrivalent VLP platform to provide protection against influenza ([240]). The approach utilizes *Nicotiana benthamiana*, an Australian relative of the tobacco plant, as an upstream bioreactor ([240], [241]). Specifically, the *S* gene from SARS-CoV-2 in its prefusion conformation is inserted into a bacterial vector (*Agrobacterium tumefaciens*) that then infects the plant cells ([240], [241]). Expression of the S glycoprotein causes the production of VLPs composed of S trimers anchored in a lipid envelope that accumulate between the plasma membrane and the cell wall of the plant cell ([241]). Because these VLPs do not contain the SARS-CoV-2 genome, they offer similar advantages to whole-virus vaccines while mitigating the risks ([240], [241]).

In the phase I study, 180 Canadian adults ages 18 to 55 years old were administered Covifenz as two doses, 21 days apart, with three different dosages evaluated ([241]). This study reported that when the VLPs were administered with AS03, an oil-in-water emulsion containing α-tocopherol and squalene ([242]), as an adjuvant, the vaccine elicited an nAb response that was significantly (approximately 10 times) higher than that in convalescent sera ([241]). The phase III trial examined 24,141 adults assigned to the treatment and control conditions at a 1:1 ratio between March and September of 2021 ([243]).

Covifenz was reported to be 71% effective in preventing COVID-19 in the per-protocol analysis ([243]). Efficacy was only slightly lower in the intention-to-treat group at 69%, with the VE for the prevention of moderate-to-severe disease in this group estimated at 78.8% ([243]). Over 24,000 participants were included in the safety analysis, which reported that 92.3% of vaccine recipients reported local AEs compared to 45.5% of placebo recipients, with rates for systemic AEs at 87.3% and 65.0%, respectively ([243]). The adverse effects reported were generally mild to moderate, with the most common adverse effects being injection site pain, headache, myalgia, fatigue, and general discomfort ([243]). Only three patients (two in the vaccine group) reported grade 4 events, all after the second dose ([243]). The vaccine was approved for use in adults ages 18 to 65 by Health Canada in February 2022 ([244]).

Plant-based expression systems such as the one used in Covifenz are relatively new ([241](#)) but are likely to offer unparalleled feasibility at scale given the speed and low-cost associated with the platform ([245](#)). Additionally, the Covifenz vaccine offers the advantage of being stored at 2 to 8°C. However, the worldwide footprint of Covifenz, and of VLP-based technologies against SARS-CoV-2 broadly, remains small, with only 1 VLP vaccine approved for distribution in 1 countries (Figure [3](#)). Approval and administration of Covifenz in countries outside of Canada has been limited by concerns at the WHO about ties between Medicago and the tobacco industry ([238](#), [246](#)). While other species of plants have been explored as the upstream bioreactors for plant-derived VLPs, the specific species of tobacco used increased yield dramatically ([247](#)). In December 2022, tobacco industry investors in Medicago divested, opening new possibilities for the distribution of the vaccine ([248](#)).

As a result of this limited roll-out and given that the phase III results were published only in May 2022, little is known about the real-world performance of Covifenz. However, it should be noted that the Covifenz trials were conducted in 2021, at a time during which the B.1.617.2 (Delta) and P.1 (Gamma) variants were predominant ([243](#)). Genomic analysis of 122 out of 176 cases (165 in the per-protocol population) revealed that none of the COVID-19 cases reported were caused by the original Wuhan strain ([243](#)). Instead, 45.9% of cases were identified as the Delta variant, 43.4% as Gamma, 4.9% as Alpha, and 5.8% as VOIs ([243](#)). Therefore, Covifenz and Nuvaxovid, despite both being designed based on the index strain, were tested under circumstances where different VOC were dominant, and differences in the Spike proteins of different VOC relative to the index strain could affect vaccine efficacy. As of late 2022, Covifenz has not been authorized as a booster in Canada ([249](#)), and no studies on booster doses had been released by Medicago ([250](#)).

## 0.6.5 Subunit Vaccine Summary

Subunit vaccine technology is one of the best-represented platforms among COVID-19 vaccine candidates. Development programs are underway in many countries around the world, including low- and middle-income countries ([187](#)). To date, data about the effect of viral evolution on the effectiveness of subunit vaccines has been limited. Because these vaccines were developed using the Spike protein from the index strain ([191](#), [243](#)), a potential concern has been that these vaccines could lose effectiveness against SARS-CoV-2 containing mutations in the Spike protein. Comparison of studies across vaccines suggests that some VOC, such as Alpha, may have minimal impact on vaccine efficacy/effectiveness ([251](#)). Additionally, to the extent that data is available such as from the vaccine rollout in Cuba, it suggests that real-world effectiveness remains strong against severe illness and death.

Subunit platforms offer some unique advantages. Cuba's successful vaccine development programs highlights the fact that protein subunit vaccines can be developed using low-cost technologies. Additionally, they are more feasible to store and transport ([252](#)). Hoping to build on Cuba's success and the continued lack of vaccine access in many countries, several Latin American nations have begun developing protein subunit vaccines ([253](#)).

The efficacy and effectiveness of these vaccines is also very high, especially for Nuvaxovid, Abdala, and SOBERANA 01/02/Plus, where estimates exceeded 90%. Unfortunately, there seem to be limited studies directly comparing the immunogenicity of subunit vaccines to nucleic acid vaccines, and comparing efficacies across trials is subject to bias ([254](#)). All the same, the evidence suggests that some protein subunit vaccines are able to provide extremely strong protection. Coupled with the reduced barriers to development and transportation relative to most nucleic acid vaccines, it is clear that subunit technologies are important to vaccine access.

# 0.7  Global Vaccine Status and Distribution

The unprecedented deployment of COVID-19 vaccines in under a year from the identification of SARS-CoV-2 led to a new challenge: the formation of rapid global vaccine production and distribution plans. The development of vaccines is costly and complicated, but vaccine distribution can be just as challenging. Logistical considerations such as transport, storage, equipment (e.g., syringes), the workforce to administer the vaccines, and a continual supply from the manufacturers to meet global demands all must be accounted for and vary globally due to economic, geographic, and sociopolitical reasons ([255](#)–[257](#)). As of January 12, 2023, at least 13.0 billion vaccine doses had been administered in at least 223 countries worldwide using 28 different vaccines ([63](#)).> The daily global vaccination rate at this time was 260.0 per million.

However, the distribution of these doses is not uniform around the globe. Latin America leads world vaccination rates with at least 82% of individuals in this region receiving one vaccine dose followed by the U.S. and Canada (81%), Asia-Pacific (81%), Europe (70%), the Middle East (58%), and Africa with only 33% as of November 2022 ([258](#)). It is estimated that only ~25% of individuals in low- and middle-income countries have received one vaccine dose ([8](#), [259](#)). Vaccine production and distribution varies from region to region and seems to depend on the availability of the vaccines and potentially a country's resources and wealth ([260](#)).

One effort to reduce these disparities is COVAX, a multilateral initiative as part of the Access to COVID-19 Tools (ACT) Accelerator coordinated by the WHO, Gavi, the Vaccine Alliance, and the Coalition for Epidemic Preparedness Innovations (CEPI), the latter two of which are supported by the Bill and Melinda Gates Foundation. Their intention is to accelerate the development of COVID-19 vaccines, diagnostics, and therapeutics and to ensure the equitable distribution of vaccines to low- and middle-income countries ([261](#), [262](#)). COVAX invested in several vaccine programs to ensure they would have access to successful vaccine candidates ([263](#)). However, the initiative has been less successful than was initially hoped due to less participation from high-income countries than was required for COVAX to meet its goals ([264](#)).

Additionally, the vaccine technologies available differ widely around the globe. As we review elsewhere ([6](#)), wealthier nations have invested heavily in mRNA and DNA vaccines. In contrast, as we describe above, many countries outside of Europe and North America have developed highly effective vaccines using more traditional approaches. There is a clear relationship between a country's gross domestic product (GDP) and its access to these cutting-edge types of vaccines (Figure [4](#)). Whole-virus and subunit vaccine development programs are responsible for a much higher percentage of the vaccinated populous in lower-income countries. Therefore, vaccine development programs that utilized established vaccine technologies have played a critical role in providing protection against SARS-CoV at the global level.

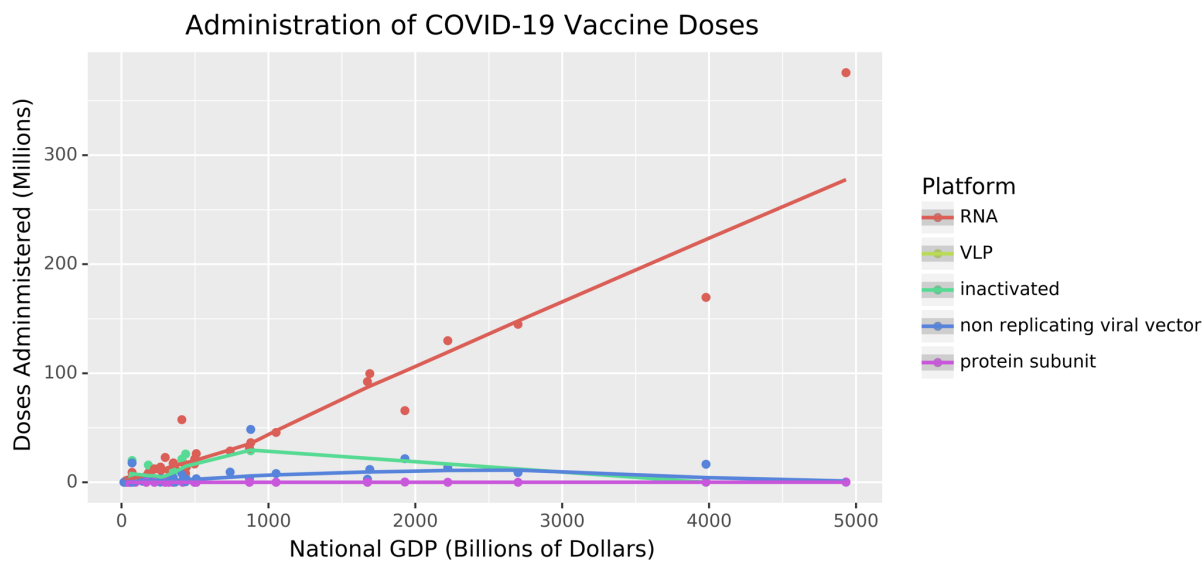

*Figure 4:* **Vaccine Distribution across Platform Types as a Function of GDP.** *The total number of doses of the original formulation of each vaccine that were distributed within each country as of January 18, 2023, by platform type, is shown as a function of GDP. These data are retrieved from Our World in Data ([8](#), [63](#)) and plotted using the Python package plotnine ([265](#)). Lines show a general trend in the data and are drawn using geom_smooth ([266](#)). The list of countries included in the dataset is available from OWID ([267](#)). See https://greenelab.github.io/covid19-review/ for the most recent version of this figure, which is updated daily. Axes are not scaled per capita because both variables are modulated by population size.*

When vaccines first became available, the wealthy nations of North America and Europe secured most of the limited COVID-19 vaccine stocks ([268](#)). Throughout 2021, low- and middle-income countries faced steep competition with high-income countries for vaccines, and the rates of vaccination reflected this unequal distribution ([269](#)). While the wealthiest countries in these regions could compete with each other for vaccines independent of programs such as COVAX ([269](#)), other countries in these regions have faced challenges in acquiring vaccines developed by the world's wealthiest nations. Fortunately, while mRNA and DNA vaccine development programs are not widespread, vaccine development using whole-virus and subunit technologies has been undertaken worldwide. China and India, in particular, have developed several vaccines that are now widely available in these densely populated countries (see online Appendix ([41](#))).

Still, many nations, especially in Africa, are reliant on the COVAX Facility, who have promised 600 million doses to the continent ([270](#)). The COVAX plan seeks to ensure that all participating countries would be allocated vaccines in proportion to their population sizes. Once each country has received vaccine doses to account for 20% of their population, the country's risk profile will determine its place in subsequent phases of vaccine distribution. However, several limitations of this framework exist, including that the COVAX scheme seems to go against the WHO's own ethical principles of human well-being, equal respect, and global equity and that other frameworks might have been more suitable ([271](#)). Furthermore, COVAX is supposed to allow poorer countries access to affordable vaccines, but the vaccines are driven by publicly traded companies that are required to make a profit ([260](#)).

In any case, COVAX provides access to COVID-19 vaccines that may otherwise have been difficult for some countries to obtain. COVAX aimed to distribute 2 billion vaccine doses globally by the end of 2021 ([272](#)). According to Gavi, as of January 2022, COVAX had distributed over 1 billion vaccines to 144 participants of the program ([273](#)), short of its target but still a major global achievement. It is envisaged that COVAX may also receive additional donations of doses from Western nations who purchased surplus vaccines in the race to vaccinate their populations, which will be a welcome boost to the vaccination programs of low- and middle-income countries ([274](#)).

In general, deciding on the prioritization and allocation of the COVID-19 vaccines is also a challenging task due to ethical and operational considerations. Various frameworks, models, and methods have been proposed to tackle these issues with many countries, regions, or U.S. states devising their own distribution and administration plans ([275](#)–[279](#)). The majority of the distribution plans prioritized offering vaccines to key workers such as health care workers and those who are clinically vulnerable, such as the elderly, the immunocompromised, and individuals with comorbidities, before targeting the rest of the population, who are less likely to experience severe outcomes from COVID-19 ([280](#)). The availability of vaccines developed in a variety of countries using a variety of platforms is likely to work in favor of worldwide vaccine access. The initiative by Texas Children's Hospital and Baylor College of Medicine to develop Corbevax, a patent-free COVID-19 protein subunit vaccine, is an important step towards vaccine equity because the manufacturing specifications can be shared globally. Corbevax can be produced at low cost using existing technology and is now licensed to Biological E. Limited (BioE), an Indian company specializing in low-cost vaccine production ([281](#)). The vaccine has been approved for distribution in India and Botswana ([282](#)).

Logistical challenges and geographical barriers also dictate the availability of certain vaccines. Many countries have had poor availability of ultra-low temperature freezers, leading to challenges of distribution for vaccines such as mRNA vaccines that require storage at very low temperatures ([283](#)–[285](#)). Furthermore, ancillary supplies such as vaccine containers, diluents for frozen or lyophilized vaccines, disinfecting wipes, bandages, needles, syringes, sharps and biological waste disposal containers are also required, which may not be readily available in geographically isolated locations and can be bulky and expensive to ship ([283](#)). While some of these challenges in vaccine rollout in low- and middle-income countries are being addressed through COVAX ([286](#)), many issues persist worldwide ([287](#)–[289](#)). COVAX also failed to distribute its promised two billion vaccine doses on time due to multiple complications ([290](#)).

Another major challenge to global vaccine distribution is vaccine hesitancy, which the WHO has designated as a leading global health threat ([291](#)). Polling in the U.S. in January 2021 suggested that 20% of individuals were reluctant to receive a vaccine at that time, with a further 31% expressing some hesitancy to a lesser extent ([292](#), [293](#)). A survey of 8,243 long-term healthcare workers in November 2020 (Indiana, USA) reported that only 69% of respondents would ever consider receiving an FDA-approved vaccine due to their perceived risk of side effects (70%), health concerns (34%), efficacy (20%), and religious beliefs (12%) ([294](#)). Notably, almost a third of parents surveyed in the United States in March 2021 expressed concerns about vaccinating their children against COVID-19 ([295]). Indeed, vaccine hesitancy has been reported as a significant barrier to vaccine distribution in countries in North and South America, Europe, Asia, and Africa ([296]–[300]). Various factors have been associated with increased vaccine hesitancy including access to compelling misinformation via social media ([301], [302]), religious and conservative political beliefs ([303]–[306]), and safety and efficacy concerns ([295]), to highlight a few. Many of the concerns regarding safety and efficacy have focused on the novel mRNA technologies due to the perceived speed of their development and expedited clinical trial process ([307]); however, general vaccine hesitancy relating to traditional vaccine platforms existed long before the pandemic and the distribution of the novel mRNA vaccines ([308], [309]). While in the United States, it was hoped that Novavax's Nuvaxovid would appeal to the vaccine hesitant ([310], [311]), but this protein subunit vaccine has not led to the uptake hoped ([312], [313]).

Overall, the vaccine landscape remains heterogeneous even as the pandemic nears its third year, with certain vaccines much more accessible in high-income countries than in low- and middle-income countries. The vaccines described in this manuscript, which were developed using well-established technologies, have played a crucial role in improving the feasibility and accessibility of vaccination programs worldwide. While the novel technologies have received the bulk of public attention in countries like the U.S., these more traditional vaccine platforms also provide safe and highly effective protection against SARS-CoV-2. Although companies developing cutting-edge technologies, namely Moderna and Pfizer/BioNTech, reported very high efficacies greater than 90% in their clinical trials ([6]), the efficacies identified in whole-virus and subunit trials have also been very high. While the clinical trial efficacy estimates for IWV and subunit have been lower, some of these trials have also reported efficacies over 80% (e.g., Novavax's Nuvaxovid with 89.7% ([194]) or Sinovac's CoronaVac with 83.5% ([71])). Variation among studies investigating the efficacy of these vaccines, especially CoronaVac, clearly indicate that clinical trials of the same vaccine might not identify the same efficacy, depending on conditions such as the specific variants circulating in a clinical trial population during the trial period. Additionally, there are many cohort- and population-level characteristics that can introduce bias within and between clinical trials ([254], [314]), and the extent to which these different factors are present may influence trial outcomes. While head-to-head comparisons of VE across different studies may therefore not be appropriate, the results make it clear that effective vaccines have been developed with a wide variety of technologies. The vaccines discussed here, which took advantage of well-established approaches, have proven to be especially valuable in pursuing vaccine equity.

## 0.8 Conclusions

Much attention has focused on the most novel vaccine technologies that have been deployed against SARS-CoV-2, but the established vaccine platforms discussed here have all made a significant impact on human health during the twentieth century and in some cases even earlier. The COVID-19 pandemic has demonstrated new potential in these established technologies. In the early 2000s, these technologies were explored for managing SARS-CoV-1 ([315], [316]), but the epidemic was controlled before those efforts came to fruition ([317]). Similarly, these technologies were explored for MERS-CoV, but outbreaks were sporadic and difficult to predict, making vaccine testing and the development of a vaccination strategy difficult ([318]). However, in the COVID-19 pandemic, most of these technologies have been used to accelerate vaccine development programs worldwide. Therefore, they are also offering the opportunity to respond quickly to an emergent pathogen.

While these tried-and-true technologies do not always produce vaccines with the highly desirable VE reported in mRNA clinical trials (which exceeded 90%), the efficacies are still very high, and these vaccines are extremely effective at preventing severe illness and death. Furthermore, some vaccine development programs using established technologies, especially protein subunit vaccines, have seen remarkably high VE and vaccine effectiveness. Some protein subunit vaccine phase III trials generated VE estimates of over 90%, comparable to those in the mRNA vaccine trials. Additionally, in some cases, such as Cuba's highly successful vaccine development program, these vaccines have been developed by and for low- and middle-income nations. As a result, the greater accessibility and stability of these vaccines makes them extremely valuable for the global effort to mitigate the loss of life from SARS-CoV-2. The outcomes of the response to COVID-19 suggests that these established vaccine technologies may continue to play an important role in tackling future viral threats.

# 1   Additional Items

## 1.1  Competing Interests

| Author | Competing Interests | Last Reviewed |
|---|---|---|
| Halie M. Rando | None | 2021-01-20 |
| Ronan Lordan | None | 2020-11-03 |
| Alexandra J. Lee | None | 2020-11-09 |
| Amruta Naik | None | 2021-04-05 |
| Nils Wellhausen | None | 2020-11-03 |
| Elizabeth Sell | None | 2020-11-11 |
| Likhitha Kolla | None | 2020-11-16 |
| COVID-19 Review Consortium | None | 2021-01-16 |
| Anthony Gitter | Inventor of patent US-11410440-B2 assigned to the Wisconsin Alumni Research Foundation related to classifying activated T cells. | 2023-01-11 |
| Casey S. Greene | None | 2021-01-20 |

## 1.2  Author Contributions

| Author | Contributions |
|---|---|
| Halie M. Rando | Project Administration, Software, Visualization, Writing - Original Draft, Writing - Review & Editing |
| Ronan Lordan | Project Administration, Writing - Original Draft, Writing - Review & Editing |

| Author | Contributions |
|---|---|
| Alexandra J. Lee | Writing - Original Draft, Writing - Review & Editing |
| Amruta Naik | Writing - Original Draft, Writing - Review & Editing |
| Nils Wellhausen | Project Administration, Writing - Review & Editing |
| Elizabeth Sell | Writing - Original Draft, Writing - Review & Editing |
| Likhitha Kolla | Writing - Review & Editing |
| COVID-19 Review Consortium | Project Administration |
| Anthony Gitter | Project Administration, Writing - Original Draft, Writing - Review & Editing |
| Casey S. Greene | Conceptualization, Supervision |

## 1.3 Acknowledgements

We thank Nick DeVito for assistance with the Evidence-Based Medicine Data Lab COVID-19 TrialsTracker data. We thank Yael Evelyn Marshall who contributed writing (original draft) as well as reviewing and editing of pieces of the text but who did not formally approve the manuscript, as well as Ronnie Russell, who contributed text to and helped develop the structure of the manuscript early in the writing process and Matthias Fax who helped with writing and editing text related to diagnostics. We are also very grateful to James Fraser for suggestions about successes and limitations in the area of computational screening for drug repurposing. We are grateful to the following contributors for reviewing pieces of the text: Nadia Danilova, James Eberwine and Ipsita Krishnan.